# AI of Brain and Cognitive Sciences: From the Perspective of First Principles

**Authors**：Luyao Chen[1,2†], Zhiqiang Chen[1,3†], Longsheng Jiang[1,4†], Xiang Liu[1,5†], Linlu Xu[1,5†], Bo Zhang[1,4†], Xiaolong Zou[1,2†], Jinying Gao[1,3], Yu Zhu[1,3], Xizi Gong[2], Shan Yu[3]*, Sen Song[4]*, Liangyi Chen[5]*, Fang Fang[2]*, Si Wu[2]*, & Jia Liu[4]*

**Affiliations:**
1. AI of Brain and Cognitive Sciences Research Group, Beijing Academy of Artificial Intelligence, Beijing 100084, China
2. School of Psychology and Cognitive Sciences, Peking University, Beijing 100871, China
3. Brainnetome Center and National Laboratory of Pattern Recognition, Institute of Automation, Chinese Academy of Sciences, Beijing, 100190, China
4. Tsinghua Laboratory of Brain and Intelligence, Tsinghua University, Beijing, 100084, China
5. State Key Laboratory of Membrane Biology, Institute of Molecular Medicine, Peking-Tsinghua Center for Life Sciences, College of Future Technology, Peking University, Beijing 100871, China

*Corresponding Authors. Email: liujiaTHU@tsinghua.edu.cn; brain@baai.ac.cn

Note: The authors with '†' are equal contributors to this work. The name of the co-authors is listed in alphabetical order.

# Content

**Introduction: The principles that AI researchers need to know**

Nowadays, we have witnessed the great success of AI in various applications, including image classification, game playing, protein structure analysis, language translation, and content generation. Despite these powerful applications, there are still many tasks in our daily life that are rather simple to humans but pose great challenges to AI. These include image and language understanding, few-shot learning, abstract concepts, and low-energy cost computing. Thus, learning from the brain is still a promising way that can shed light on the development of next-generation AI.

The brain is arguably the only known intelligent machine in the universe, which is the product of evolution for animals surviving in the natural environment. At the behavior level, psychology and cognitive sciences have demonstrated that human and animal brains can execute very intelligent high-level cognitive functions, such as flexible learning, long-term memory, and decision-making in an open-ended environment. At the structure level, cognitive and computational neurosciences have unveiled that the brain has extremely complicated but elegant network forms to support its functions. Over years, people are gathering knowledge about the structure and functions of the brain, and this process is accelerating recently along with the initiation of giant brain projects worldwide. So, what is the most important knowledge AI should learn from the brain?

Here, we argue that at the current stage, the general principles of brain functions are the most valuable things to inspire the development of AI. These general principles are the standard rules of the brain extracting, representing, manipulating, and retrieving information, and they are the foundations of the brain executing other higher cognitive functions. In some sense, they are the principles guiding the running of the brain, and here we call them the first principles of the brain.

This paper collects six such first principles summarized by the research team, "AI of Brain and Cognitive Sciences", in the Beijing Academy of Artificial Intelligence (BAAI). They are attractor network, criticality, random network, sparse coding, relational memory, and perceptual learning. On each topic, we review its biological background, fundamental property, potential application to AI, and future development.

# Chapter 1 Attractor Dynamics: Canonical Models for Neural Information Processing

*Xiaolong Zou, Si Wu\**

## 1.1 Introduction: The dynamical system theory and the attractor network

The brain is composed of a huge number of neurons, which form various networks via synapses between them. It is generally believed that the computations of single neurons are rather simple, and it is the dynamics of neural networks that accomplish brain functions. Simply stated, a neural network receives inputs from the external world and other brain regions, and its state evolves to carry out information processing. Accordingly, the dynamical system theory is a valuable mathematical tool to quantify how the brain performs computation via networks.

A dynamical system describes how a set of variables evolve over time, which provides a powerful mathematic framework to investigate complex behaviors. Generally, a simple deterministic dynamical system can be described as,

$$\frac{du(t)}{dt} = f(u(t), x(t)),$$

where $u(t)$ is the state vector of the system at time $t$, e.g., the firing rates of the neural population, and $x(t)$ is the external input. $f(\cdot)$ is a non-linear function that specifies the evolution rule of the state $u$.

In a dynamical system, different state evolution rules and varied external inputs can create diverse dynamical phenomena in a dynamical system. In a recurrently connected neural network, the firing rate vector of the neural population evolves and forms a trajectory in the state space of the network. A state vector is called a stable attractor if all its neighboring states flow into it. A network with stable attractors is called an attractor network. Similarly, network states may flow into a close-loop trajectory and generate periodic responses. Such a close-loop trajectory is called a limit cycle attractor, and a network with such an attractor state is called an oscillatory attractor network. There are also other attractor dynamics, such as saddle point and chaotic attractor. These diverse attractor dynamics enable a neural system to perform various brain functions [1,2,3,4,5,6].

Here, our focus is on the attractor networks with stable attractor states, and we argue that they form the building blocks of the brain for information representation, manipulation, and retrieval. Intuitively thinking, attractors are the only states of neural networks that enable neural systems to store information reliably against noises ubiquitous in environments and the brain. The study on attractor dynamics has a long history. As early as 1972, Shun-ichi Amari studied a recurrently connected neural network and found that it could exhibit discrete attractor dynamics [7]. Such a phenomenon was rediscovered by John Hopfield and was later called the Hopfield model [8]. Hopfield demonstrated that the Hopfield model can store a pattern as a stable attractor of the network, which explains the associative memory of the brain. Continuous attractor neural networks (CANNs), in which attractors are continuously located in the state space, are an extension of discrete attractor networks. CANNs were first introduced by Shun-Ichi Amari [9] and later were successfully applied to explain orientation tuning in the visual cortex [10], head direction representation [11], and spatial navigation [12] et al. So far, discrete and continuous attractor networks have been widely used as canonical models in the literature for elucidating various brain functions.

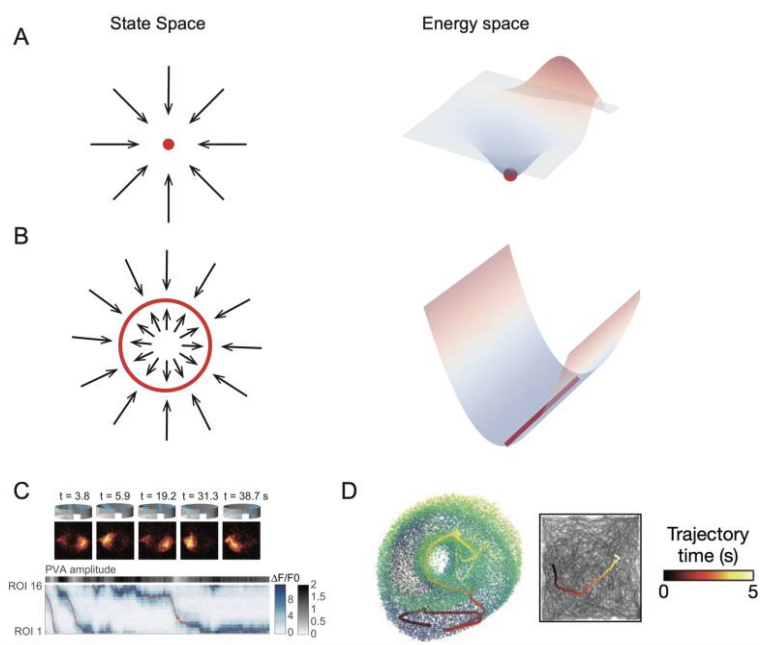

Figure 1. Different attractors and evidence for attractor dynamics. (A) Discrete attractor network. Arrows indicate how the network state evolves in the state space. An attractor state is a point in the state space, which all neighboring states flow into (left panel), corresponding to a local minimum in the energy space (right panel). (B) One-dimensional CANN. All attractor states form a ring structure in the state space (left panel), forming a smooth manifold in the energy space (right panel). (C) The one-dimensional CANN in the fly's head direction system. The activity bump of the network tracks the fly's head rotation during flight, manifesting the characteristic of a CANN. Adapted from [17]. (D) The torus-like structure of a CANN in the population activity of a single grid module in the entorhinal cortex of a rat. Adapted from [20]

## 1.2 Discrete vs. Continuous Attractor Dynamics and their biological backgrounds

### 1.2.1 Discrete vs. continuous attractor dynamics

We can roughly divide attractor networks with stable states into two types, discrete and continuous attractor networks, as shown in Fig.1A-B. In a neural network, the state vector corresponds to a point in the state space of the network that all its neighboring states evolve into it. Thus, this state vector is an attractor corresponding to a local minimum in the energy space of the network, as shown in Fig.1A. In a discrete attractor network, each attractor has its own attractive field. Starting from a random state, the recurrent dynamics of the network decrease the energy until it drives the network state to a neighboring attractor state having the local minimum of the energy. Such characteristic of discrete attractor dynamics enables the network to correct input noises and retrieve the clean memory representation. Discrete attractor networks are often used to model working memory, long-term memory, and decision-making, et al.

Unlike discrete attractor models, attractors in a CANN are allocated continuously in the state space, which forms a smooth manifold, as shown in Fig.1B. This property enables attractor states in a CANN to quickly transfer from a state to nearby states, called neutral stability. It brings many appealing computational properties of CANNs [3,4,12], such as path integration, evidence accumulation, anticipative tracking. In structure, the translation invariance of synaptic connections between neurons, i.e., the connection strength between two neurons determined by their distance instead of their preferred locations in the feature space, is the critical property of a CANN.

Notably, both discrete and continuous attractor networks are mathematically idealized models. In the natural neural system, a network structure in between discrete and continuous attractor networks is likely adopted to encode information, which has computational properties partially overlapped with both attractor networks, i.e., to retrieve information relatively reliably as discrete attractor dynamics and to change states relatively quickly as continuous attractor dynamics.

**1.2.2 Biological backgrounds**

Experimental studies have accumulated compelling evidence for the existence of attractor dynamics in the brain [13,15,16,17,18,20,21]. As early as 1971, Fuster et al. [13] found that many prefrontal neurons discharge spikes persistently during the memory retention period of a delayed response task. These persistent activities are believed to play a fundamental role in working memory and are an indication of attractor dynamics [14]. Compared with discrete attractors, continuous attractors have more predictable features, which have been found in different cortical areas, such as the hippocampus and its related areas. For example, the head direction system can integrate the head rotational velocity and information of the external cues to encode the head direction [15].

A CANN is successfully used to simulate the head-direction system, which encodes angles of head direction via continuous attractors. Recently, experimental studies [16] found that the neurons in the head direction system of the fly form exactly a ring topology structure as a one-dimensional CANN. In addition to the structure resemblance, Kim et al. [17] performed a large population recording of neurons in the head direction system of the flying fly. They found that the dynamical property of the system also agrees well with that of a CANN, e.g., a local activity bump in the system can be triggered by an external cue. It tracks the head direction smoothly as the fly head turns (Fig.1C). The bump can be maintained in the darkness as an attractor. Furthermore, when a new bump is created, it suppresses the bump at the previous location. When a stimulus is applied in the neighboring location of the bump, the bump activity will drift to the stimulus location.

The phenomena above are the properties of a one-dimensional CANN. Two-dimensional CANNs are also found in the hippocampus and entorhinal cortex of animals, such as rodents. In the hippocampus, place cells fire when animals visit some specific locations in the environment and are proposed to form an internal representation of spatial locations. They exhibit many attractor properties, such as persisting in the dark [18], which can be well-explained by a CANN [19]. In the entorhinal cortex, grid cells encode the abstract spatial locations and form a regular triangular-lattice discharge pattern. Like the head direction system, the grid cell network can integrate motion and visual cues to represent spatial location [12,19]. To account for the periodic grid pattern, a CANN predicts that the states of grid cells form a torus topologically in the state space. Recently, Gardner et al. [20] performed a large recording of grid cells during walking and sleep and confirmed that population activity of grid cells from the same module (cells that share identical periods and orientations) forms a toroidal topology (Fig.1D). Furthermore, CANNs do not only involve in spatial navigation, but also in other cognitive functions, such as evidence accumulation. For example, Mante et al. [21] found that when a monkey performs a context-dependent decision-making task, as the evidence accumulates, the population state in the monkey's prefrontal cortex evolves along a line attractor until reaching a decision threshold.

In sum, the accumulated experimental and modeling evidence suggests that attractor networks can be regarded as a canonical computational model employed by the brain for information processing.

**1.3 Information representation with attractor dynamics**

Attractor neural networks have been widely used in modeling brain functions. Here, we review their fundamental properties in information representation, which lay the foundation of their roles in cognitive functions.

**1.3.1 Robust information representation in attractor networks**

An attractor network can represent information robustly. In a discrete attractor network, the information of memory is stored as an attractor state. Given a partial or noisy cue, the network dynamic evolves into an attractor state, and the corresponding memory is retrieved. Different attractors correspond to different local energy minima and have their own basins of attraction. If noise perturbation is not too strong to push the network state to escape from the basin, the attractor state is stable. Thus, the memory information is encoded robustly. Unlike discrete attractor networks, attractors in a CANN form a flat manifold in the network state space, and they are partially robust to noise. If noise perturbation is orthogonal to the manifold, the network state is stable by the attractor dynamics. However, if noise perturbation is along the manifold, the network state will diffuse on the attractor manifold, moving from one attractor to the neighborhood. To prevent memory drift in a CANN, Lim et al. [22] proposed a mechanism relying on negative derivative feedback to retain the stability of the attractor state.

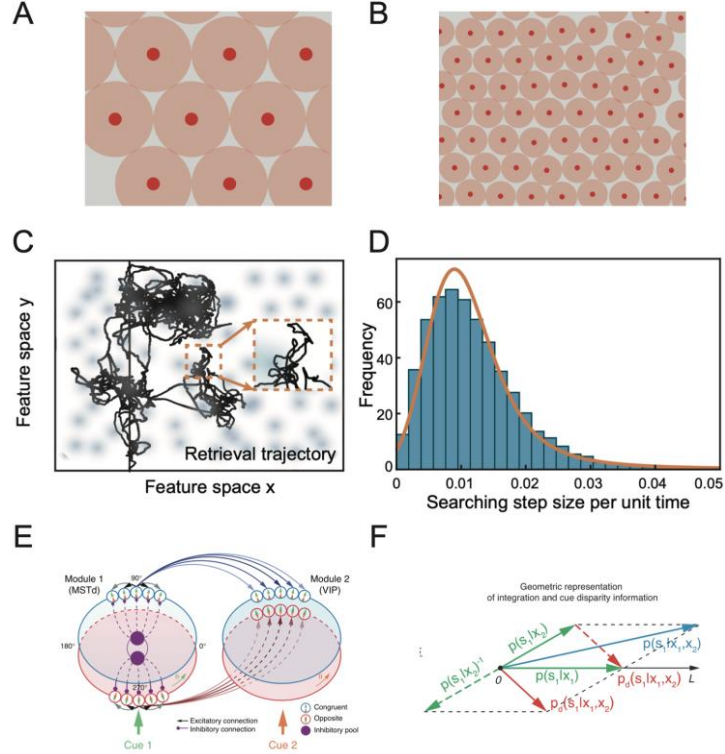

Figure 2. Information representation in attractor neural networks. (A-B) The tradeoff between capacity and robustness in discrete attractor networks. (A) When a small number of patterns are stored in a fixed-size network, each attractor (small red circle) has a large attractive basin (large light-red circle) and can tolerate a larger amount of noise. (B) In contrast, storing a large number of patterns decreases the attractive basin of each attractor and weakens each attractor's noise robustness. Adapted from [6]. (C-D) Information search in a CANN. (C) An example of an information search in the attractor space. The retrieval trajectory exhibits a characteristic of Levy flights. (D) The histogram of the searching step size follows a power-law-tailed distribution. Adapted from [25]. (E-F) Information integration in reciprocally connected CANNs. (E) The network consists of two modules. Each module contains two CANNs, one consisting of congruent neurons and the other of opposite neurons. (F) A geometric interpretation of information integration and segregation is implemented by the reciprocally connected CANNs. Adapted from [27]

### 1.3.2 Memory capacity of attractor networks

Memory capacity refers to the number of memories that can be stored reliably in an attractor network. Several factors affect the memory capacity of an attractor network. One is noise. When too many memories are stored in a network, the attraction basin of each attractor shrinks, which decreases the noise tolerance of the attractor, as shown in Fig.2A-B. Another factor is memory correlation. When memory patterns are highly correlated, they will interfere with each other and destabilize memory retrieval. For a Hopfield model of N neurons, its memory capacity is about 0.14N, if all memories are random patterns. However, when patterns are strongly correlated, neighboring attractors merge into one, which introduces incorrect representation, called spurious attractors. Furthermore, when the number of learned memory patterns exceeds the capacity of a Hopfield model, the stability of all existing attractors will be destroyed. To increase the memory capacity of attractor networks, many approaches have been proposed, from learning rules to network structures, such as novelty based Hebbian rule [23] and modular Hopfield network [24].

**1.3.3 Information search in attractor networks**

In addition to large memory capacity, a good information processing system also requires efficient information search. In an attractor network, memories are usually retrieved in a content-addressable manner, i.e., the network performs similarity computation via attractor dynamics and retrieves the memory most similar to the cue. In a large-capacity network, finding the correct one among massive attractors is challenging. For example, in a free memory retrieval task, participants need to search and recall animal names as many as possible. A good recall strategy is a suitable combination of local memory search and a large jump in the memory space, manifesting the Levy flight behavior. Dong et al. [25] demonstrated that in a CANN with noisy neural adaptation, the dynamics of the network state shows interleaved local Brownian motion and long-jump motion, exhibiting the optimal information searching behavior as Levy flight, as shown in Fig.2C-D.

**1.3.4 Information integration between attractor networks**

Attractor networks can also interact with each other to accomplish information integration between modalities. Recently, Zhang et al. studied how reciprocally connected CANNs can realize multisensory information processing [26,27], as shown in Fig.2E-F. In their model, they consider that each module contains two group of neurons, each of which forms a CANN, and their tuning functions are either congruent or opposite with respect to the modality input. They showed that coupled CANNs with congruent neurons implement information integration, while coupled CANNs with opposite neurons implement information segregation, and the interplay between them achieves concurrent multisensory integration and segregation efficiently. This study demonstrates that interconnected attractor networks can support information communication between cortical regions.

Limited by space, we have only introduced some fundamental properties of attractor networks. In application, when extra elements are induced in the network structure, the attractor network can exhibit richer dynamical behaviors with associated appealing computational properties. For example, a CANN with spike frequency adaptation (SFA) can perform anticipative tracking [28], a CANN with feedback connections can exhibit oscillatory tracking behaviors [29], and a CANN with noisy SFA can implement efficient sampling-based Bayesian inference [30].

## 1.4 Conclusion: Global Neuronal Workspace Theory

In past years, limited by data and technology, computational modeling in the field has been mainly focused on studying the dynamics and functions of single neurons and local circuits. Recently, boosted by technical advances and giant brain projects worldwide, a large amount of data about the details of brain structure and neural activity is emerging. It is a timely request to build large-scale network models to simulate higher cognitive functions. Attractor networks, as canonical models for neural information processing, serve as the building blocks for us to carry out this task.

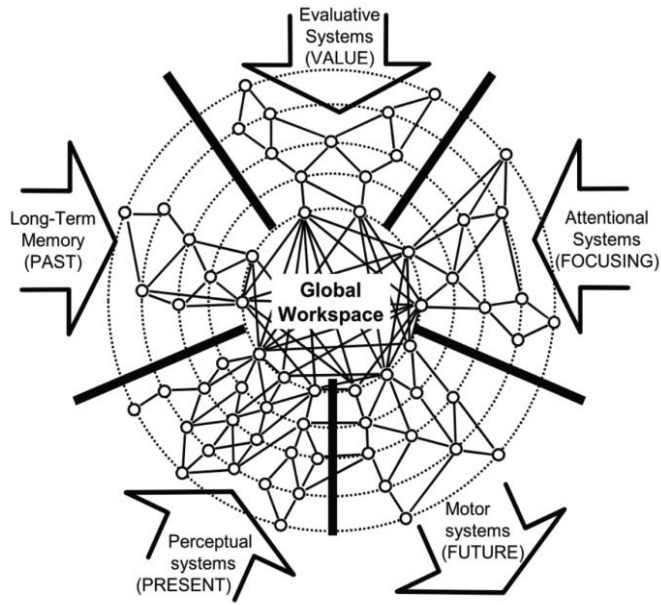

Figure 3. The GNW hypothesis. The model contains a global, shared processing module and many local, specialized processing modules. [31]

Here, we use the global neuronal workspace theory (GNW) [31] as an example to discuss the potential applications of attractor networks. GNW proposes a framework illustrating how the brain achieves consciousness. According to GNW, the brain is divided into a shared, global processing module and many distributed, specialized processing modules, as shown in Fig.3. Each unique module processes information from one modality, such as the visual, auditory, olfactory, or motor system. In contrast, the global module receives and integrates information from all specialized modules, and meanwhile broadcasts the integrated information back to those local modalities. To achieve this goal, an abstract information representation interface is required, which allows different modules to communicate with each other. In this sense, CANNs, which have been demonstrated as canonical models in both experiments and theoretical studies, naturally serve as the unified framework to represent, transform, integrate, and broadcast information between modules. It will be interesting to investigate this issue in the future work.

# Chapter 2 Criticality: Bringing New Perspectives to the Brain and Artificial Intelligence

*Zhiqiang Chen, Jinying Gao, Yu Zhu, Shan Yu\**

## 2.1 Introduction: Brain dynamics

The framework of criticality is a powerful tool to understand and analyze complex systems because many systems in physics and nature are in a critical state. In the past 20 years, researchers found that biological neural networks in the brain operate close to a critical state, which provides a new perspective on studying brain dynamics. It is known that the critical state is important for brain activities/functions because it optimizes numerous aspects of information transmission, storage, and processing. In addition, some brain diseases are believed to be related to the deviation from the critical state, which also opens a new window for diagnosing and treating these diseases. In the field of artificial intelligence, the framework of the critical state is used to analyze and guide both the structural design and weight initialization of deep neural networks, suggesting that operating close to a critical state may be considered one of the fundamental principles governing computations in neural networks.

## 2.2 The critical state and its main characteristics

In statistical physics, a homogeneous state in a material system with identical physical and chemical properties is called a phase [1]. For example, water can be in the solid phase, liquid phase, or gas phase. When temperature changes, water can change from one phase to another, which is called a phase transition [2,3,4]. The critical state is a type of so-called second-order phase transition, which indicates that the system underlies the transition from an ordered phase to a disordered phase. At the edge between order and disorder, or the "edge of chaos", the critical state exhibits many special properties.

### 2.2.1 Phase transition and critical state

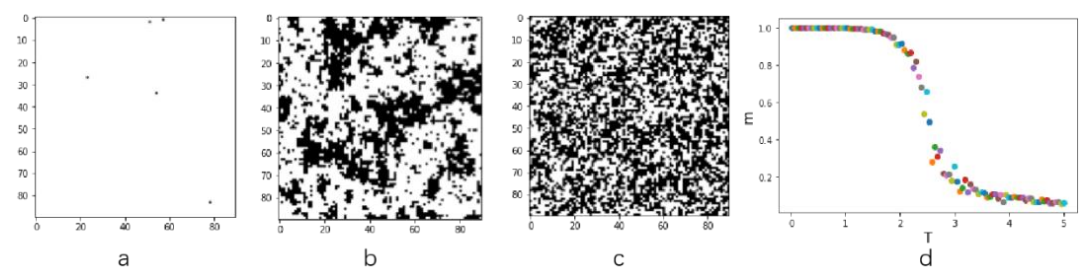

Figure 4. A Monte Carlo simulation of the Ising model [7]. It shows spins for each lattice at (a) low temperature, (b) critical temperature, and (c) high temperature, respectively. White means pointing up and black means pointing down. (d) The average magnetic moment $m$ over temperature T. Each colored point means a different temperature. (Color is only for display and has no special meaning.) All figures are produced by code from link[1].

Fig. 4 shows the phase transition process and critical state of ferromagnetic materials by simulation of the Ising model [5,6,7,8]. In the Ising model, the competition of spin interaction and thermal motion causes the ordered and disordered phases. Fig. 1a and 1c show the ordered phase at low temperatures and the disordered phase at high temperatures, respectively. When the temperature changes from low to high as shown in Fig. 1d, the system will undergo a phase transition. At the temperature on the edge of phase transition, as shown in Fig. 1b, order and disorder are in equilibrium, and neither can dominate the entire system. At this temperature, the system is extremely complex and is in a so-called critical state. In the ordered or disordered phases, the domain size [9] distribution is concentrated at larger or smaller sizes, respectively. But in the critical state, the domain size distributes at almost all sizes. And the distributions at different scales are self-similar, which means that the distributions are fractal [10] and scale-free [11]. This self-similar distribution can be mathematically formalized as a power law [12]: $p(x) \propto x^{-\alpha}$. If we use the log-log coordinate system, the distribution will appear as a straight line. Power-law distribution is an important characteristic of the critical state. When a system is in a critical state, many statistical variables obey the power-law distribution.

### 2.2.2 Self-organized critical state

In addition to precisely adjusting the temperature in the Ising model or control parameters in other systems to reach the critical state, some systems can spontaneously reach the critical state, which is called self-organized criticality (SOC). A well-known SOC model is the sandpile model [13,14]. In this model, sand is slowly dropped to a surface and the slope of the pile gradually increases. When it reaches a critical slope, adding more sand will make the pile collapse, forming variously sized sand avalanches that make sand leave the pile, thereby returning to a critical slope. The size of these avalanches obeys the power-law distribution. Since SOC was proposed, it has been used to explain many complex phenomena, including fluctuations in the economic system [15], voting in elections [16], pulsar [17], black hole [18], and so on. Importantly, more and more studies have found that the brain might also be in a self-organized critical state [19,20]. In the next chapter, we will introduce the related studies in detail.

### 2.3 Criticality in the brain

Over the last two decades, by recording neuronal activity in either brain tissues cultured *in vitro* or the intact brain *in vivo*, many experiments showed that cortical networks can also self-organize into a critical state and the spatial and temporal distributions of cascading neuronal activity approximately obey the power law. This phenomenon is called "neuronal avalanches", which provides us with a new perspective to understand the dynamics of brain networks. It not only has numerous computational advantages in information processing but also opens new windows for the treatment and diagnosis of diseases caused by the deviation from the critical state.

---

[1] https://zhuanlan.zhihu.com/p/436304252

**2.3.1 Neuronal avalanches**

In the cortical network, each neuron receives input from a large number of surrounding neurons through synaptic connections. When the input reaches its threshold, an action potential will be generated, which will be transmitted to other neurons, causing other neurons to fire. Beggs and Plenz [21] discovered the commonalities between the biological neural network and the sandpile model and confirmed the conjecture of the critical brain for the first time with multi-electrode array recordings in brain tissues. They found that both sizes and lifetimes of neuronal avalanches obey a power law, which is an important characteristic of the critical state. Later, other researchers recorded neuronal activity in different cerebral cortices of different species, in both awake and anesthetized states that reconfirmed the characteristics of a power-law distribution of neuronal avalanches in spontaneous activities of the network. It shows that operating close to a critical state of brain networks is a general feature, and the balance of excitability and inhibition plays a key role in helping maintain the critical state [22]. To examine if neuronal avalanches are a unique phenomenon of spontaneous neuronal activity, Yu et al. [23] recorded extracellular unit activity and the local field potential in monkey's premotor and prefrontal cortex during motor and cognitive tasks, which showed that the network activities involved in active information processing are also maintained close to a critical state, suggesting that neuronal avalanches are unified phenomena of neuronal activity both during the rest and behaving states.

**2.3.2 Evidence beyond power law**

Power-law distribution is strong evidence for the critical state [24]. However, power law alone cannot be considered the definition of criticality. Many systems that are obviously not in critical states can also generate power-law distributions [25]. Fortunately, the existence of other evidence beyond power-law distribution can corroborate the identification of the criticality. If we tune the temperature in the Ising model or "control parameters" in other critical systems, the power-law distribution will rapidly disappear, and a phase transition will take place [24]. While power-law distributions unrelated to the critical state will not undergo a phase transition [24], in the nervous system, the balance between excitation and inhibition can serve as a control parameter. When blocking either excitatory or inhibitory transmitters with drugs, the changing of the E-I balance breaks the original power-law distribution and presents two different phases (i.e., subcritical and supercritical, respectively) [22]. Furthermore, many fractal features appear when the system is in a critical state. When scaling avalanche shapes of different durations, a clean collapse occurs, as would be expected in a critical state [26]. Such evidence from multiple perspectives nicely converges and gives strong support that the brain is indeed organized close to a critical state.

**2.3.3 Computational advantages of criticality**

Why is the brain in a critical state? Further studies showed that a network in a critical state has obvious advantages in information transmission, storage, and processing. These advantages have been confirmed both in critical models and biological experiments. In 2006, Kinouchi et al. [27] constructed a simple neural network that constrained the local branching ratio of each neuron to be 1, at which the network was considered to reach a critical state. Through modeling, it is found that the representation of the input by the network has the optimal dynamic range under this branching ratio. When the branching ratio is set to be less than 1, the network is considered to be in a subcritical state, in which the network does not distinguish weak inputs clearly. When the branching ratio is set to be greater than 1, the network is considered to be in a supercritical state, in which the network will soon reach saturation. Later, Shew et al. [22] grew slice cultures on the surface of multi-electrode arrays and conditioned the culture environment to reach criticality, thereby demonstrating that systems in criticality can most sensitively perceive signals with an amplitude spanning several orders of magnitude. In addition, other researchers have confirmed that systems in critical conditions have optimized computational capabilities [28], maximum mnemonic repertoire size [29], and maximum fidelity of information transmission [30]. Hu et al. [31] demonstrate that working memory (WM) has the optimal performance and the maximal sensitivity/flexibility in a biologically plausible WM model [32] when the network is regulated to the critical state by dopamine.

**2.3.4 Diseases caused by the deviation of the critical state**

Large deviations from the critical state in the brain may cause epilepsy [33], burst suppression [34], and schizophrenia [35], all of which manifest as a disruption of the critical state and thus interfere with the normal transmission or processing of information. The rapid development of computational models has allowed researchers to use the critical state framework to model the above-mentioned diseases and thus provide a new perspective to understand the complex experimentally recorded data of the nervous system. Dean et al. [36] developed a system that can track the trajectory of the seizure through bifurcation space, which can be used as a means of long-term prediction and diagnosis of epileptic seizures. Xin et al. [37] applied critical dynamics analyses on resting-state functional magnetic resonance imaging data from major depression patients. They found that the macroscopic brain network of severely depressed patients deviates from a critical state and maintains a subcritical state. Importantly, electroconvulsive therapy (ECT) restores the critical state, accompanied by alleviation of the depression, which provides a mechanistic explanation for both severe depression itself and the therapeutic effects of the ECT. Although still exploratory, these studies open a new window for the diagnosis and treatment of neurological diseases [38].

## 2.4 Applications of the critical state in artificial intelligence

Given the successful application of critical states in the analysis of many complex systems including the nervous system, some researchers have also tried to use critical states to study artificial neural networks, for example, to improve reservoir computing and empower deep neural networks.

### 2.4.1 Improving reservoir computing

Reservoir computing (RC) typically refers to a special computational framework of recurrent neural networks (RNN) in which the trainable parameters only reside in the final read-out layer, namely the non-recurrent output layer, while all other parameters are randomly initialized and fixed in the subsequent computation [39,40,41,42] (see also Chapter 3 Random Network). Currently, RC models have been successfully applied to many computational problems for tasks such as temporal pattern classification, recognition, prediction, and action sequence control [43,44]. RC models work well only when the network is in a critical state, sometimes also called the "echo state" [39]. It means that we need to carefully initialize the network connections [45,46]. Inspired by the short-term synaptic plasticity (STP) in biological neural networks, Zeng et al. [45] implemented a self-originated criticality (SOC) scheme induced by short-term depression (STD) in the RC model, which automatically adjusts the state of the RNN close to criticality. STD greatly strengthens the robustness of the neural network so that it can adapt to long-term synaptic changes while maintaining optimal performance endowed by criticality. It also suggests the potential mechanisms that the brain employs to organize plasticity at different time scales, thus maintaining an optimal state (critical state) of information processing while allowing for the internal structural changes required for learning and memory [45].

### 2.4.2 Empowering deep neural networks

Deep neural networks have achieved great success compared to shallow networks. To theoretically explain this phenomenon, Poole et al. [47] combined Riemannian geometry with the mean-field theory of high-dimensional chaos to reveal that transient changes in the chaotic state (critical state) are the origin of the exponential expressivity with increasing depth in deep stochastic networks. In addition, they demonstrated that this property is absent in shallow networks. This finding has significant implications for the structural design of networks and provided a theoretical base for the superior performance of existing deep neural networks. Motivated by this work, Schoenholz et al. [48] developed a mean-field model for the gradient of deep networks and showed that the deep networks could be well-trained only when it remains at or near a critical state. The ordered and chaotic phases established by Poole et al. [47] correspond precisely to the regions where the gradients vanish and explode. Theoretical analyses showed that weight initialization on the edge of chaos, i.e., the input-output Jacobian mean squared singular value of a deep network should remain close to 1, which leads to a substantial increase in learning speeds [49]. Furthermore, Pennington et al. [50,51] adopted free probability theory to analyze the entire distribution of computed input-output Jacobian singular values, taking depth, random initialization, and nonlinear activation functions as independent variables. However, none of the above studies touched on learning rules. Oprisa et al. [52] found that the activation frequency of neurons does not follow the power-law distribution, and the critical state does not appear spontaneously in classical deep neural networks. They pointed out that designing learning rules to induce critical states in a network is a fundamental missing part. In addition, Farrell et al. [53] also found that strongly chaotic RNNs (super-criticality) appear better adept at learning the balance of expansion and compression and thus achieve better performance. Therefore, it's still an open issue whether being in the critical state is always helpful.

## 2.5 Conclusion: The future study on criticality

The critical state gives us a new perspective to study biological and artificial neural networks. At present, the framework of criticality has not only been used to understand neural dynamics and brain diseases but also to analyze the operation of deep neural networks and guide the design for further improvement. Through theoretical analysis and numerical simulations, we know that the critical state of the network can be controlled by some simple control parameters, such as branching ratio, spectral radius, and input-output Jacobian singular values. This makes it possible to analyze or tune the overall behavior of complex networks through statistics that are easy to observe. We believe that the framework of criticality will play an even more important role in helping us better understand the constraints applied to artificial neural networks and to design better architectures as well as dynamical rules to improve its performance in complex information processing.

# Chapter 3 Random Network: A Potential Foundation for Information Encoding

*Longsheng Jiang, Sen Song\**

## 3.1 Introduction: Dimensionality

Since Hubel and Wiesel's groundbreaking finding of neuron tuning to bar orientations [1], neurophysiologists thrive on finding neurons that possess clean tuning curves to single specific stimulus features. In this endeavor, however, neurophysiologists often find themselves in a confusing situation where many observed neurons simultaneously reflect different features nonlinearly [2]. These neurons are said to have nonlinear mixed selectivity [2,3,4,5,6,7]. The effort to understand it has ushered a shift to a population-based analysis of neurons [8,9,10]. Population coding represents neural responses in a high-dimensional state space, in which the activities of each neuron represent one dimension of the space [see also Chapter 1 Attractor Dynamics]. In this neural space, more information is discriminately decodable [10].

The presence of nonlinear mixed selectivity neurons increases the dimensionality of the representations. As a result, the otherwise linear inseparable representations become linearly separable [3] and can be further processed by the downstream structure of the brain [11,12]. To ensure high dimensionality, mixed selectivity also needs to be diverse [3]. To investigate circuits in the brain that support diverse mixing while remaining simple, some researchers suggest random networks may be at work [13,14,15,16,17,18,19]. In random networks, the synaptic weights of neuron connections are randomly sampled from some distributions. These connections mix signals as inputs to downstream neurons. The inputs go through nonlinear mapping within the neurons. In this way, the neurons are endowed with nonlinear mixed selectivity. The randomness of the connections endows the diversity of mixing. Growing biological evidence aligns with this notion [20,21,22,23,24,25,26,27,28,29,30,31,32].

## 3.2 Biologically observed random connectivity

As early as 1888, Cajal [33] discovered two different structural features of a neuron cell: the axon, which courses straight through the neuropil, and the dendrites, which extend widely as if they try to contact as many axons passing through as possible [34]. The complex networks of interlacing neurons with such a basic structure, packed in limited space in a brain region, suggest a distributed circuit model for neural computation. In this model, a neuron sends signals to and receives signals from as many other neurons as possible [35]. Ideally, the distributed circuit model prescribes full connectivity between neurons at the physical limit. While full connectivity is biologically implausible [36], it is possible to use random connectivity to approximate full connectivity [35].

Such randomness has been observed in animals' brains, where the eminent evidence of random connectivity comes from olfactory systems. The random connections are from the odorant-specific glomeruli in the antennal lobe to the Kenyon cells (KCs) in the mushroom body [20]. Such connections defy any anatomic and functional structures in the connections. Even for the two hemispheres of the same fly's brain, the connection structures are different [21]. A model of KC implementing random connectivity and other biological constraints recapitulates the actual KC responses [22]. Likewise, in the olfactory cortex of rodents, the connections from the olfactory bulb to the piriform cortex (PCx) are random [23]. A random network model of PCx largely recapitulates the pairwise relationships between odor representations.

Another typical example is the cerebellum [24,25,26]. The pathway from mossy fibers, through granule cells, to Purkinje cells involves random connections. The mossy fibers conduct signals into the cerebellum, which connect randomly to the ubiquitous granule cells. Then, through parallel fibers, the granule cells connect to widely extended synapses of Purkinje cells. A Purkinje cell serves as the readout neuron in the cerebellum.

Although less consensual [27], strongly supportive evidence comes from visual systems. The functional connections from the cones in the retina of macaque monkeys to retinal ganglion cells (RGCs) are almost random for receptive field centers and exactly random for receptive field surround [28]. Down the visual pathway, randomness is the feature of the connections between the RGCs and relay cells in the lateral geniculate nucleus (LGN) for cats. A model introducing random connectivity between the RGCs and relay cells in the LGN matches experimental data and facilitates the interpolation of visual space [30]. In addition, the connections from the LGN to the primary visual cortex (V1) of cats may be random as well [31], as a model with random connectivity gave rise to orientation maps, and the characteristics of the modeled receptive fields matches experimental observations [31]. For the animals such as rodents in which orientation maps do not exist, orientation selectivity can emerge from random connectivity between the LGN neurons and the V1 neurons [17]. This is because the spatial offsets of the receptive fields of the ON and OFF cells converge at a cortical neuron. The simulated offsets generated by the model match those observed in experiments.

Besides sensory systems, random connectivity may also exist in the circuits for action [13], decision-making [32], and navigation [15]. A randomly connected feedforward network simulated the connections between the premotor cortex and the primary motor cortex (M1) [13]. This network provided a rare example of replicating both the regularity and heterogeneity of neural responses in M1. The heterogeneous element contributed to decoding electromyographies (EMGs) from M1 neuronal activities. A recurrent network with random connection weights was used to simulate the posterior parietal cortex (PPC) in an evidence accumulation task for decision-making [32]. The simple model reproduced the experimental data in the distribution of single-neuron selectivity, patterns of pairwise correlations, etc. In simulating the spatial memory network of fruit flies, the random network can support persistent, continuous attractors [15], and thus can memorize locations.

### 3.3 Divergent network architecture

If random networks indeed mean to increase the representational dimensionality, the architecture of the networks should reflect this purpose. The maximum dimensions of a neural space allowed by a neuron population are the number of total neurons. For increasing dimensionality, at the extreme of complete random connectivity, the population of post-connection neurons should be larger than that of pre-connection neurons. Therefore, the networks should be in a divergent architecture [26].

The divergent architecture is indeed present in biological brains [37,38]. The antennal lobe to mushroom body pathway in fruit flies provides a clear example. The antennal lobe contains about 50 glomeruli. A fraction of about 150 project neurons innervate the glomeruli and project to about 2500 KCs in the mushroom body [39]. The olfactory bulb of rodents hosts about 1800 glomeruli [40], whereas the downstream PCx hosts millions of neurons [37]. In the human visual pathway, the optic nerve from 1 million relay cells in LGN projects to the order of 100 million cortex neurons in V1 [27]. In the rat's cerebellum, about 7000 mossy fibers are estimated to expand to about 209,000 granule cells presynaptic to one Purkinje cell (i.e., the readout neurons) [41]. These phenomena suggest that creating rich high-dimensional representations through divergent networks with random weights might be an underlying computational principle. Indeed, this hint has been encountered in artificial intelligence (AI) since the early 1990s [42].

**3.4 Random networks in artificial intelligence:**

Random networks in AI refer to artificial neural networks (ANNs) whose partial weights are initialized randomly and are not adaptive during training. Researchers were initially attracted to these networks either because they are easy to set up for analysis [43,44], or because their training is much faster [45]. However, researchers soon found random networks performed surprisingly well; their testing accuracy was close to fully trained models [46], in applications including short-term forecasting, image recognition, and biomedical classification [42]. Motivated by these observations, researchers studied the characteristics of various random networks.

Two categories of networks are extensively investigated. Feedforward networks and reservoir-like recurrent networks [see also Chapter 2 Criticality]. In the feedforward networks, the input neurons connect with random weights to a hidden layer with a much larger size. In reservoir computing, the input neurons connect to a reservoir of inner neurons that have random connections with each other. Examples of feedforward networks include the random vector functional link network (RVFL) [47], radial basis functional link network [48], feedforward network with random weights [49], no-prop algorithm [50], weight agnostic network [51], and random CNN [52]. Examples in reservoir computing include echo state network (ESN) [53], liquid state machine (LSM) [54], and deep ESN [55] (See [56] for detailed descriptions).

All these models share three features: (1) the hidden layer, or the reservoir, creates a high dimensional representation of the inputs [57], (2) the weights of connections to the output neurons need to be linearly optimized [42], and (3) the network performance is robust to different realizations of the random weights [46]. What can be concluded from these observations is that the architectures of trained ANNs, rather than the fine-tuned connection weights, are more responsible for task performance. A further interesting study shows that even the architecture itself can be random. Architectures created by random graph generators show good classification accuracy on the ImageNet (79% for randomly wired networks versus 77% for ResNet-50) [58]. These observations suggest that randomness is hardly a naïve trick, but it may be fundamental to machine intelligence. This point echoes similar conjectures in neuroscience, as we summarized above. The effectiveness and efficiency of the random network and its potential to embody an undiscovered computational principle thus motivates many works to study it analytically [5,37,41,59,60].

**3.5 Computational theories about random networks:**

The purpose of the random network, in biological brains or computers, is to increase representational dimensionality. But what exactly is the benefit of high dimensionality? For $I$ different points, the probability of these points to become linearly separable in a $d$ dimensional representation space is [61]:

$$\Pr(d) = \left(\frac{1}{2}\right)^I \binom{I-1}{d-1}.$$

As $d$ increases, the probability of separation also increases, making the downstream decoding easier. To harvest the decoding advantage, the dimensionality should increase rather rapidly.

To elucidate the increment of the representational dimensionality, here we focused on a simple feedforward network. Suppose the hidden layer has $M$ neurons, which create a $M$ dimensional hidden space. The dimensionality $d$ of representations in the hidden layer is defined as [62]:

$$d = \frac{(\sum_{i=1}^{M} \lambda_i)^2}{\sum_{i=1}^{M} \lambda_i^2},$$

where $\lambda_i$ are the eigenvalues of the covariance matrix of the points in the hidden space. According to this definition, if each dimension is independent and has the same variance, then $d = M$. However, if dimensions are correlated, then $d < M$ [41].

The dimensionality $d$ is constrained by the network's architecture. Suppose the input layer has $N$ neurons and the representation in the input layer is $N$ dimension. Also, suppose each hidden neuron connects to $K$ input neurons and the connection weights are homogeneous. When $N$ and $M$ are large, the dimensionality $c$ of mixed inputs arriving at the hidden layer, before passing through the nonlinear activation function, is [41]:

$$c \approx \frac{N}{1 + \frac{N}{M} + \frac{(K-1)^2}{N}}.$$

For a fixed $K$, a divergent architecture with $M \gg N$ increases the value of $c$. As seen, c is guaranteed $c \leq N$, as linear mixing cannot exceed the original dimensionality. However, after nonlinear activation, the representational dimensionality $d$ can be $N \leq d \leq M$ [41].

Random connectivity between the input and hidden layers can further expedite the dimension increase. If the connection weights are sampled from a distribution with mean $\mu$ and variance $\sigma^2$, and when $M \gg N \gg 1$, the dimensionality $c$ of mixed inputs is [41]:

$$c \approx \frac{N}{1 + \frac{N}{M} + \frac{(K-1)^2}{N}\left(\frac{\mu^2}{\mu^2 + \sigma^2}\right)^2}.$$

For fixed $N$, $M$, $K$, and $\mu$, a higher variance $\sigma^2$ results in a larger $c$, and correspondingly a larger representational dimensionality $d$. Analytically, heterogeneous connection weights facilitate dimension increase. The heterogeneity can be obtained either through training or random sampling. The above-reviewed networks in neuroscience and in AI show that randomly sampled weights suffice. A following question is whether random networks can be universal for implementing any functionality.

The universality of random networks has been mathematically proved [60, 63]. Igelnik and Pao proved that under proper nonlinear activation functions and ranges of random weights, any sufficiently smooth functions could be approximated with an error of order $O(1/\sqrt{M})$ [57]. Rahimi and Recht [60] provided probabilistic proof for a rich family of functions. It states that with at least probability $1 - 2\delta$, the random networks' approximation error is bounded by the order $O(\sqrt{\log(1/\delta)}/\sqrt{M})$ [60]. In either proof, it is shown that any general function can be approximated by random networks to any level by increasing the size $M$ of the hidden layer.

It is equally important to know what algorithm random networks are actually implemented. Dasgupta et al. found that the mechanism of the olfactory circuit of fruit flies is very similar to locality-sensitive random hashing algorithms in computer science [59]. That is, the locality-sensitive hashing and the olfactory circuit preserve the similarity between representations [64]. Researchers found that adding random hashing to deep learning significantly increases the training efficiency (using 5% of the computation while keeping within 1% of the accuracy) [65]. It is therefore argued that locality-sensitive hashing might be a computational principle in the brain.

### 3.6 Limitation of random network

The biological evidence, the engineering practice, and the theoretical analysis seem to point to a view that a distributed random network is enough for achieving cognitive functions. However, this view is overly simplified. Random networks must couple with other network characteristics to achieve sophisticated functions. These characteristics include convergent readout [66], plasticity [67,68], excitatory-inhibitory balance [14,44], and sparseness [5,14,37].

In neural circuits, the divergent randomly connected layer is often followed by convergent connections to readout neurons [22,38]. The convergent structure is necessary for maintaining the inter-individual consistency of neural representations [66]. Since the connections are initialized randomly for individual brains, two brains hardly have the same connectivity pattern. As a result, inter-individual consistency of representations is lost at the neuron level. That is, the same stimuli activate different neurons in individuals. However, inter-individual consistency still exists at the population level. Converging the activities from the population to a readout neuron inherits consistency [66].

It is widely recognized that the convergent connections after the random layer are plastic. They are experience-dependent [20] and require training [42]. Plasticity is critical for inter-individual consistency. After two realizations of random networks have their convergent connections trained on only one stimulus with Hebbian learning, the realizations display consistent generalization over the whole set of stimuli [68]. Furthermore, it was suggested that the random connections of the divergent layers may undergo Hebbian learning [67]. After adding this simple mechanism to the random connections, the results from the model simulation surprisingly match the experimental selectivity profiles of a neuron population [67].

Besides convergently connecting to readout neurons, another salient feature of the divergent random layer is that it is subject to feedback inhibition, in the olfactory systems in fruit flies [22] and rodents [23], and in the cerebellum [26]. One possible function of inhibition is to ensure the selectivity of neurons. In a distributed network where a neuron connects to thousands of other neurons, the fluctuations of the summed inputs are overwhelmed if all the connections are excitatory [44]. Inhibition activities balance the excitatory activities so that the expected summed activities are 0 [68]. Fluctuations, therefore, are highlighted, and the network exhibits high stimulus selectivity [14].

Another possible function of inhibition is to create sparse representations [see also Chapter 4 Sparse Coding] in the divergent layer [5], which may implement a winner-take-all algorithm [59]. A higher level of inhibition gives rise to sparser representations. Sparseness is critical as it controls the trade-off between discrimination and generalization of the network [5]. Discrimination in the high dimensional state space created by the divergent layer requires differentiating the points, whereas generalization requires grouping the points. Analysis of computational models showed that high sparseness results in lower discriminability but higher generality [5]. In fact, sparseness is important to neural circuits as it may be by itself a biological computation principle [69,70].

All the additional features, including convergent readout, plasticity, excitatory-inhibitory balance, and sparseness, are indispensable to neural circuits. They must be built on top of the divergent random layer, and random connections are the prerequisite.

## 3.7 Conclusion: Beyond dimensionality and sparseness

Random networks are the simplest neural circuits that produce mixed selectivity commonly observed in neurophysiology. While countering the common sense that functionality can only arise from organized networks, random networks have been found in various systems in biological brains over the last decades. Meanwhile, randomness has been employed in AI as a computationally efficient method for building ANNs. Due to their peculiarity and effectiveness, random networks have attracted many theoretical studies, which help reveal the underlying principles.

The principles can be explained on three conceptual levels [71]. On the computational level, random networks are universal function approximators like trained neural networks. With a divergent architecture, random networks create high dimensional state space, in which the discriminative decoding is more flexible and achievable. On the algorithmic level, random networks work as locality-sensitive hashing algorithms in computer science. These algorithms can greatly save computational requirements for training deep networks. On the implementation level, random networks are the most plausible physical realization for distributed networks in the densely packed neuropil in the brain.

The principles of random networks should not be viewed in exclusive light. It is only fully functional when working with other features, including convergent readout, plasticity, excitatory-inhibitory balance, and sparseness. Therefore, random networks should not be viewed or applied as isolated circuits.

The studies over the last decade have helped us realize the importance of random networks and clarify some crucial concepts. However, more questions still need to be answered. On the computational level, despite dimensionality and sparseness, we hardly know more about the representations in random networks. What does the manifold of intrinsic states look like in the state space? On the algorithmic level, the distributions for random sampling weights are still empirical and arbitrary. How should the distributions be specified? Should prior knowledge be used? Should the generated weights be fixed or go through slow Hebbian learning? On the implementation level, the brain also possesses modularity, e.g., functional columns. How should modularity reconcile with the randomly distributed network structures? In answering these questions, our knowledge of random networks will be advanced. Up then, we may indeed confirm that random networks represent a fundamental principle for intelligence.

# Chapter 4 Sparse Coding: Its Unique Features and Potential Advantages in the Brain

*Xiang Liu, Linlu Xu, Liangyi Chen\**

## 4.1 Introduction: Sparsity in information processing

The brain is a machine that stores and processes information. To achieve these functions, accurate quantification and reasonable representation of external information are needed [1]. Sparse coding strategies act as a critical way to achieve these goals. The brain exploits sparsity mechanisms at multiple levels, including the visual, olfactory, tactile, and other perceptual levels, which participate in processes such as cortical information processing [2]. Discussion of these mechanisms is essential to understand the principles of neural system organization and intelligence formation.

## 4.2 Advantages of sparse coding

Sparse coding means that the number of firing neurons is only a tiny fraction of the total number of neurons at any given time [3]. Thus, sparsity is a relative concept, and there is no clear threshold. The advantages offered by a sparse code are best appreciated when compared with two more extreme coding schemes: local coding and dense coding [4]. Local coding is sometimes called one-hot code: each neuron is involved in coding one item only, as in the case of 'grandmother cells', and there is no overlap between representations of any two items. In another extreme, dense coding is a fully distributed coding: each item will be represented by the combined activities of all neurons within a population [see also Chapter 3 Random Network]. Sparse coding is in between and often enjoys the advantages of both sides [5].

Sparse coding offers a good trade-off between encoding capacity (as well as energy efficiency) and decoding difficulty. Since no overlap is allowed in local coding, a population of N binary neurons can maximally represent N different items. More energy will be consumed as more neurons must be recruited to encode more items. Therefore, the energy available to the brain sets the upper limit for the local coding. On the other hand, dense coding dramatically improves the representational ability by allowing N binary neurons to encode $2^N$ items. Even if only a few (up to K) neurons may be activated simultaneously for an item, as in the case of sparse coding, the total number of different codes will be $\sum_{k=1}^{K} \binom{N}{k}$. Compared to local coding, this configuration saves neurons and consumes much less energy to encode equal pieces of information. However, the readout of a distributed code is not trivial and can be challenging to learn in a biologically plausible manner. In contrast, the association of a local code and its output can be established using a simple Hebbian mechanism; thus, learning becomes more efficient if the neural activity patterns are sparsely coded [3, 6].

Sparse coding also balances generalization and interference. In local coding, each pattern is orthogonal to any one of the other patterns. As there is zero similarity between different patterns, it is impossible to generalize associations from one pattern to another. Dense and sparse codings allow partial overlap and different levels of similarity between codes, enabling generalization between items with similar codes. However, dense coding dictates that many items (up to 50% of all items if the code space is fully occupied) may activate one neuron. This broad tuning may lead to interferences between different patterns [7]. Forming a new association between a code pattern and an output unit with learning may interfere with old memories associated with overlapping codes and shared connection weights [5]. Sparse coding may help address such catastrophic forgetting [8] and minimize interference between patterns [9]. In the extreme, local coding does not suffer from interferences, and multiple items can be simultaneously represented.

Finally, sparse coding explicitly represents natural stimuli's structure that sharply tunes the neuronal responses. The receptive fields resemble the frequent structures encountered in the environment, so that the activation of only a few neurons can represent a natural stimulus. Combining with overcomplete basis, sparse coding may produce a piecewise flattened representation of the curved manifold on which natural stimuli clustered , simplifying the representation and analysis in later stages [3]. These advantages support more efficient encoding, transmission, and storage of information by organisms.

**4.3 Sparse coding in the brain**

Sparse coding is ubiquitous in the neural system, especially in primary sensory cortices. Previous studies reported sparse coding related to the visual, auditory, olfactory, and somatosensory systems [2,10,11,12]. In addition, it is also found in the brain areas associated with movement and memory [13,14,15]. Some views on sparse coding in neural systems are that sparse coding is embodied in three forms: population sparsity, lifetime sparsity, and connection sparsity [16,17]. Population sparsity means that the number of neurons firing at a given moment in a population is sparse. Lifetime sparsity describes neurons that sparsely fire during their lifetimes [18]. Previously, it was speculated from the perspective of metabolic energy constraints that, at a given time, about 0.5% to 2% of neurons within the cerebral cortex are capable of simultaneous activation [19,20]. In 2018, Tang et al. directly measured the sparsity of neuronal population activation in the awake primate cerebral cortex. They presented a large number of natural scene pictures to macaques and simultaneously recorded neural activity in the superficial layers of the macaque primary visual cortex using large-field two-photon microscopy. While only 0.5% of these neurons responded to arbitrary natural pictures, each neuron responded strongly to less than 0.5% of all natural images [12]. As V1 neurons in macaques show high population and lifetime sparsity, the high selectivity of neurons may perform sparse coding in the macaque visual cortex.

The biological brain also exhibits connection sparseness, often measured by the percentage of neurons in a group or a layer connecting to another layer. According to multi-electrode intracellular recording, pyramidal neurons have lower local interconnectivity between layers than within. For example, pyramidal neurons are inter-connected within one layer at the ratio of 1:10 [21] or 1:4 [22], while this ratio jumps to 1:86 [23] or 1:29 [22] between layers. This connectivity ratio varies between species and sublayers and differs slightly between experimental measurements, but the overall properties are consistent. Interestingly, Carl Holmgren et al. found a low connectivity ratio between pyramidal neurons within the same sublayer, and there is a significant decrease in connectivity between neurons as the distance between them increases. This contrasts with the high connectivity ratio between interneurons and pyramidal neurons, which is not altered by increasing their physical distances [24]. Therefore, pyramidal neurons may adopt a closely full connection to the interneurons while not following the same pattern when connecting with each other. Recent advances in electron microscopy (EM) allow for the precise reconstruction of all synapses within a volume of the cortex, which provides a gold standard for establishing the neuronal connectome. Using large-volume serial EM, Wildenberg et al. found that cortical neural networks are sparser in primates than in mice, presumably constrained by the costs of synaptic maintenance [25].

Many theories and models have been proposed regarding how sparse coding is implemented in natural neural systems. Rozell and Olshausen [26] proposed a Locally Competitive Algorithm, which attributes sparse coding as the result of mutual inhibition between neurons from the same region. People believe this mechanism may relate to compressed sensing theory [27]. Sparse coding in the olfactory system possesses a similar mechanism with an extra feedback cortex [28], and such a mechanism also appears in the retina. Sparse coding and sparse neural activities enable efficient information compression or dimension reduction. Ganguli pointed out that random projection can project a high-dimensional sparse signal to a low dimension space without disrupting the structure of the original signals [27]. These random projections can be implemented simply as a random synaptic connectivity matrix in a neural system [see also Chapter 3 Random Network]. In this way, the neurobiological implementation of sparse coding signals may be trivial and universal.

**4.4 Theoretical study and application of sparse coding**

Sparsity and sparse coding have received long-standing attention. Barlow proposed his efficient coding hypothesis in 1961 [29], followed by the suggestion that sparsity may be an essential principle for perceptual representation [30]. Other work showed that natural images can be sparsely coded and that such coding properties are very similar to neuronal cell responses in the V1 [31].

Many people have tried to understand and explain sparsity's underlying mechanisms and related biological significance. To help machine learning and intelligent algorithm developments, people have explored the advantages of sparsity and sparse coding from several aspects, including but not limited to coding ability, robustness and generalization, compressed sensing, and information transfer efficiency [1,27,32]. These studies have led to the development of dictionary learning algorithms [33] and new algorithms like Hierarchical Temporal Memory (HTM) [32] that utilize sparsity for neural computation.

### 4.4.1 Sparse coding improves representation efficiently

What is the advantage of sparse coding? From a learning perspective, representing and encoding external perceptual information in the brain should be a process of extracting "valid information." To address this point, Ma et al. [34] proposed *The Principle of Parsimony* [1]: an intelligent system tries to extract low-dimensional information from external information and organize it into a compact (i.e., efficient) and structured form. The extracted low-dimensional information should obey three characteristics: compressibility, linearity, and sparsity. Sparse coding can effectively improve the representation capability of the coding system. Ma et al. used the quantity *rate reduction* to describe the representation capability to illustrate this point. Using methods such as sparse dictionary learning [33] or independent variable analysis, features under sparse coding can be guaranteed to be as orthogonal as possible, which maximizes the representation capability of different features.

### 4.4.2 Sparse effectively improves the performance of machine learning models

Sparsity has also been applied in machine learning. Sparsity can be used to counteract dimensional catastrophe [27]. Theoretically, for learning in a single-layer perceptron (for a simplified model of a single neuron), minor generalization errors only arise when the number of a training set is larger than the number of synapses (i.e., the number of weights) in the perceptron [35]. This is unacceptable because in complex models such as deep learning networks the number of parameters is much larger than the number of samples. Lage-Castellanos et al. [36] find that introducing L1 regularization in a single-layer perceptron can evade the above problem. Therefore, when sparsity is added to the model as prior knowledge, it can improve the performance of the machine learning model.

In fact, a single neuron is much more complicated than a perceptron. By mimicking the properties of cortical pyramidal neurons, Jeff Hawkins et al. [32] developed a learning model based on the sparse distribution of neurons, HTM, which have a hierarchical synaptic structure and a distal synaptic reception. This structure appears to affect the transmission of information between the layers because the distal synapses do not necessarily transmit the neural emissions to the cytosol. However, their study demonstrated that with sparse coding (i.e., the number of nerves in the lower layers is much larger than the number of activated neurons), the model is robust to input noise while ensuring matching accuracy [32]. The robustness is precisely the result of the properties of high-dimensional sparse vectors.

### 4.4.3 Sparsity and compressed sensing

Since Tao et al. [37,38] proposed and developed the theory of compressed sensing in 2006, the application of this theory in neuroscience and artificial intelligence has also been extensively studied [26,27,39,40]. Compressed sensing theory states that information sparsely encoded in high dimensions can be transmitted through a low-dimensional channel without information loss if certain conditions are satisfied. This may play important roles in the brain, since the signals the neural system receives from the outside world have certain sparsity (natural images, sounds, smells, etc.). The efficiency and accuracy of information transmission to deeper brain regions are fundamental problems, and compressed sensing theory may provide a solution. For example, researchers have proposed a theoretical model that is feasible in neural systems for long-range brain communication [27,40], which can compress high-dimensional information in long-range brain transmission, improving information transfer efficiency in neural systems. Compressed sensing theory can also enhance the working memory of biological systems [39,41]. Theoretical models have demonstrated that neural circuits can record sparse signals with lengths exceeding the number of their own cells. In contrast, they can only record signals with lengths not exceeding the number of neurons.

**4.5 Conclusion: Sparsity and dimensionality**

As Suryaz et al. [27] point out, "The problem of storing, communicating, and processing high-dimensional neural activity patterns, or external stimuli, presents a fundamental challenge to any neural system.", processing and learning from external information is an essential task of the neural system. Moreover, high-dimensional information is often sparse in nature. Sparse encoding strategies may be a necessary and feasible approach used by biological brains to process external information and can improve the efficiency and robustness of this procession.

**Chapter 5 Relational Memory: Neural Population Coding and Manifold**

*Bo Zhang, Jia Liu\**

**5.1 Introduction: The relational memory**

Knowledge storage, one of our indispensable cognitive abilities, has been a critical issue of neuroscience and computational science. As to the critical issue of how massive amounts of daily experiences are organized and eventually stored as knowledge by the brain, recent neuroscience discoveries indicate that the brain's memory system uses the frame of reference, through which relations of different information are precisely formed in the Medial Temporal Lobe (MTL). Nowadays, the concept of the reference frame has reconciled the experimental observations of hippocampus function in both non-spatial and spatial memory, and it is driving a new and promising research direction: relational memory. We here propose the reference frame as one of the first principles of neuronal population.

**5.2 Long-standing debates on MTL function**

The MTL consists of the hippocampal formation and several anatomically adjacent structures, including the perirhinal cortex, entorhinal cortex, and parahippocampal cortex [1]. The hippocampus is the earliest known area that is essential for memory from patient studies. Two famous cases of studies were the patient H.M. [2] and patient R.B. [3]. Patient H.M. underwent bilateral medial temporal lobe resection after uncontrollable seizures. The operation sharply reduced the incidence and severity of seizures without gross changes in personality and general intelligence. However, a grave loss of short-term memory was unexpectedly observed. H.M. could no longer recognize the hospital staff, find his way to the bathroom, and recall the day-to-day events of his hospital life. In contrast, his earlier memory before his admission to the hospital remained vivid and intact. Similar to H.M., patient R.B. had short-term memory impairment following amnesia but with intact long-term memory and cognitive abilities, and histological examination confined his memory loss to the bilateral lesion of the CA1 field of the hippocampus. Those clinical cases of memory defect, around the 1960s, suggested the distinction between short- and long-term memory and highlighted the function of the MTL in memory consolidation, that is, lesions to the hippocampus cause failure in the transformation of learned experiences from short-term memory into long-term memory. With the development, both short- and long-term memory were categorized as declarative memory, and the relevant earlier studies deepened our understanding of MTL function and promoted the construction of classic memory theory [4]. However, spatial memory, another category of memory, has long been neglected by the theory of memory.

  Remarkable progress in the understanding of the spatial coding of the brain began in the 1970s, O'Keefe and his colleagues [5] implanted electrodes in the dorsal hippocampus (field CA1) of rats, and the activities of total eight units were monitored during rat behaviors to auditory, visual, olfactory, and tactile stimuli in a 24 cm × 36 cm platform. As a result, 8 units showed responses solely when the rat was situated in particular locations in the platform. As confirmed by extended observations [6], 26 out of 50 units recorded from CA1 of the hippocampus were classified as place units, those units exhibited preferential response when the rat was located at a particular place in a 38 cm × 15 cm platform with three radiating arms, the selective responses were not appeared due to any sensory stimuli, rats' behavior, or any motivational factors occurring in represented locations, but depending on its locations on the platform. Those findings suggested that place cells in the hippocampus provide independent codes of location. A spatial reference map could thus be formed by place cell population, which acts as a cognitive map system [7]. The discovery of place cells challenges the traditional view of the hippocampus, and results in a contrasting view: are cells in the CA1 of the hippocampus really specialized for events (non-spatial information, e.g., the identity of hospital staff, the way to the bathroom, or the day-to-day experiences) or are they specialized to code spatial locations (spatial information, e.g., I am standing near the door of the room)? Although researchers, represented by Howard B. Eichenbaum, began to propose a bridging framework for the hippocampus's function as a relational processing system [8], the research of declarative memory and spatial memory has been primarily conducted in parallel although both of them focused on the hippocampal function (Fig. 1a).

**5.3 Universal code of MTL system: the reference frame**

Moser and his colleagues [9] implanted electrodes to the dorsocaudal medial entorhinal cortex (dMEC) of rats, with the goal of seeking evidence of map-like structural organization for place cell formation in upstream areas of the hippocampus, they recorded 57 dMEC neurons when 11 rats collect scattered randomly thrown pieces of food in square box. As a result, the striking spatial organization appeared from the firing fields of recorded neurons, the firing fields concentrate at the vertices of a grid of equilateral triangles that regularly tessellate the whole environment. They named the neurons "grid cell". The repeating receptive field of grid cells was initially hypothesized to encode the animals' location and direction of movement on a path-integration-based map. Lately, it became the clue to build the bridge between declarative memory and spatial memory by Constantinescu and her colleagues [10]. They hypothesized that the so-called map is organized conceptually by grid cells and tested if humans use a grid-like symmetric code when navigating in an abstract rather than physical space. In their experiment [10], a conceptual 2D "bird space" was built with the lengths of the bird's neck and leg varying in continuous dimensions, and each bird stimulus was associated with a Christmas symbol. Participants morph birds according to the ratio of neck-length change over leg-length change by keypress in MRI scanning after extensive training of the task. In this case, the ratio varied corresponding to the movement direction in the bird space. The logic behind the design is that if the bird space information was coded by grid cells, the direction-selectivity (alignment or misalignment of grid cell axes) would show a 6-fold modulation after participants move in different directions. As expected, the stable grid-like signal was revealed by the mean activity within the MEC as a function of morphing directions in bird space.

Constantinescu et al.'s finding, as a milestone, showed the first evidence of the grid cell that codes not only the physical space but also the conceptual events (Fig.1b). Following the "bird space", non-spatial coding of grid cell was further confirmed by various types of conceptual spaces, including the vision [11], odor [12], reward [13], and sequence [14] space. As a whole, those studies support the relational processing theory proposed by Eichenbaum, that is, the MTL exhibits a universal code that organizes non-spatial knowledge into a reference frame where knowledge could be stored or retrieved via a global relational manner (e.g., Christmas Tree is associated with the bird with a longer neck and short leg in bird space in bird space [10]), analogous to organizing spatial relations in physical space. So far, empirical evidence has reconciled the distinction between spatial and non-spatial memories by the clue of grid cell research, while the discovery of grid cells depended on the research of place cells. In 2014, the Nobel Prize in Physiology or Medicine was awarded to John O'Keefe, May-Britt Moser and Edvard I. Moser, in recognition of their discoveries of place cells and grid cells that code the spatial coordinate system of the brain. More importantly, those cells provided remarkable insights into how daily experiences from the external world are organized by the reference frame irrelevant to the categories of memory [15].

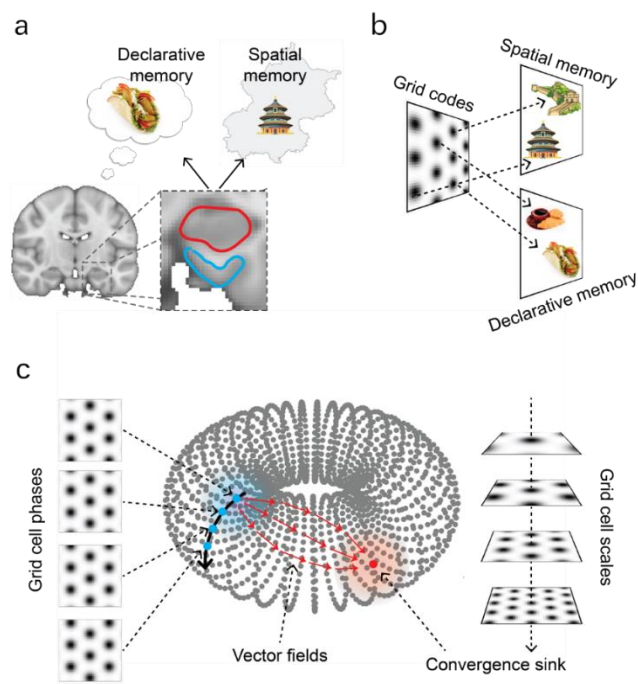

Figure 5. a) The anatomical location of the hippocampus (red outline) and entorhinal cortex (blue outline) of the MTL that codes both declarative memory and spatial memory. b) The hexagonal fields of grid cells in the entorhinal cortex serve as the reference frame that tessellates the knowledge space. c) The geometrical structure (torus) of grid cell population, in which a grid cell with a unique phase was represented by each location of the torus (left column). Blue and red shadow areas represent the start (filled blue circle) and goal location (fill red circle), both locations are coded by the unique activity strength determined by the multi-scale grid cell population (right column). The stronger activity at the goal location (convergence sink) forms the activity gradient that drives path integration via vector fields (red arrows).

The first successful implementation of the unifying framework of the MTL was recently achieved by Whittington and colleagues in 2020 [16], they built the reference frame system named as Tolman-Eichenbaum Machine (TEM). With the assumption that different aspects of knowledge are represented separately and can be flexibly recombined to represent novel experiences, the TEM machine uses the "factorization and conjunction" approach for the structural generalization of knowledge so that a broad range of neural representation observed in spatial and non-spatial memory tasks can be learned and predicted. Specifically, the task requires TEM to move on a 2D graph which was constituted by a set of nodes with each associated with a sensory observation (e.g., an image of a banana), and TEM machine that was trained by a generative model was able to predict the next sensory observation and successfully generalized the structural knowledge. As a result, the simulation of MTL's function using relational knowledge is proven to be efficient by TEM's performance and the successful replication of biological observations, including the emergence of place and grid cells, and the remapping phenomena.

**5.4 Populational coding of knowledge storage and retrieval using reference frame**

A reference frame could theoretically store the knowledge as much as possible in a continuous and quantitative manner in correspondence to a specific feature [18]. Inevitably, a single dimension of reference frame would be incapable of storing the knowledge constituted by multi-features, which has been highlighted to be the basis of two forms of flexible memory expression [17], the "transitivity" denoting the ability to judge the knowledge pairs that share a common feature (e.g., A > B & B > C, then A > C), and the "symmetry" denoting the ability associating the knowledge pairs presented in the reverse of event order (e.g., B to C, then C to B). To achieve the demand for diversity, populational coding must be required to link pieces of knowledge from multiple dimensions, which rely on hundreds of thousands of reference frames [18].

Recently, studies from neuroscience and computational science focused on the topological structure of firing rates of neurons, and their results showed evidence that neuronal population coordinates knowledge embedded in reference frames (i.e., knowledge storage) and assists with navigation in complex environments (i.e., knowledge retrieval). In view of the dynamics of populational coding, those studies shed light on the direction of revealing the principle of how neurons efficiently interact with each other for the storage and retrieval of relational memory, which as a classic issue remains to be answered for decades.

The study on the mechanism of neural interaction was initially from the field of computational modeling, like Amari's proposal of the competition-cooperation mechanism [19], which assumes a periodic weight function that forms a series of attractors [see also Chapter 1 Attractor Network]. In the scene of spatial navigation, each attractor was centered at a location of two-dimensional Euclidean space, and neural interaction is then determined by either excitation (positive) or inhibition (negative) weight of periodic function. The attractor networks, in particular one of its variants termed as "continuous attractor neural network" (CANN), have successfully simulated the pattern dynamics of place cells [20] and grid cells [21,22] so that spatial knowledge could be stably coded by the cells. On the basis of CANNs, Samsonovich and McNaughton [20] further proposed that the firing rates of grid cell population rely on a topological structure termed as "torus" (Fig.1c). In contrast to the Euclidean space where movement would be eventually bounded, a toroidal structure has no boundary to adapt to the periodic pattern of grid cells, and therefore movement starting from any location on torus would never be bounded but eventually returns to the origin. Recently, Moser and his colleagues provided the first physiological evidence which proved the existence of such a toroidal structure [23]. In their experiment, hundreds of grid cells were simultaneously recorded by high-site-count Neuropixels silicon probes when rats were engaged in foraging behavior or sleeping. After performing dimensionality reduction on the firing rates of 149 grid cells within a single scale and transforming the principal components into a 3D visualization, a torus-like structure was clearly revealed and remains stable across the animal's cognitive state (i.e., awake or sleep). In toroidal structure, the grid cell population was organized as a whole, with the inner and outer circle of the torus corresponding to the horizontal and vertical axis of Euclidean space, respectively. Therefore, each location on the torus represents a grid cell with a unique phase pair (corresponding to the inner and outer circles, respectively) that seamlessly codes physical space under a given spatial module. Given the topological morphology, the hypothesis that hundreds of thousands of reference frames were flexibly organized by high dimensional topological space beyond Euclidean space is thus directly proved.

In view of dynamics, the movement on the surface of the toroidal structure corresponds to the process of a series of knowledge retrieval. However, what is the principle that each movement determines the next "location" (e.g., A to B then B to C)? Inspired by the physiological evidence of grid cells, the gradient of module size (spacing of field sizes) was found along the dorsal to the ventral part of the dMEC, and grid cell phases were randomly distributed, the modularity system of the brain was proposed [9, 24]. The system illustrates the mapping relationship between phase and position so that animals' position could be uniquely specified by a set of phases,

$$\vec{\mathcal{X}} = (x \bmod \lambda_1, x \bmod \lambda_2, \ldots x \bmod \lambda_N),$$

where the symbol $mod$ represents modulo operation, $x$ and $\lambda$ denotes current location and gradient of module size, respectively. As a result, the grid cell population was theoretically proved to have the capacity to code maximum number of locations up to $(\prod_{\alpha=1}^{N} \lambda_\alpha)$-1 in a space [24]. On the basis of the modularity system, the winner-take-all mechanism, proposed by Bicanski and Burgess [25], assists in forming the vector from the current location to the next location towards the goal relying on the phase of the grid cell with maximal activation out of the population across modules.

Evidence supporting this algorithm was recently found by O'Keefe and his colleague [28]. In their experiment, 456 CA1 place cells were recorded from 5 rats while rats navigated to a goal for food reward in a honeycomb maze. Surprisingly, the firing patterns of 142 out of the 456 cells showed vector fields converging on a location near the goal. When summing up the vector fields across cells, maximal firing rates (termed as "convergence sink") could be clearly seen from the population vector map when the animal was oriented towards the goal (Fig.1c). In addition, the vector fields flexibly adapted to the location of goal, that is, the maximal firing rates from populational coding re-organized towards to new goal when the location of the original goal was shifted. Those results directly supported the winner-take-all mechanism that the ERC-HPC system creates a vector-based model to support flexible navigation.

Following the recent understanding of the populational coding of neurons on vector-based navigation, neural networks in the field of machine learning have shown great success in brain-inspired applications for recognition memory [26] and spatial navigation [27]. Bicanski and Burgess developed a visual memory model that uses the grid cell population to drive saccades so that familiar faces, objects, and scenes could be recognized [26]. In the navigation part of the model, each location of gaze on the image was represented by a unique vector of the firing rate from 100 grid cells with nine spatial scales at that location of gaze. Given both the current and a random goal location, the displacement vector that determined the saccade could be computed using the winner-take-all mechanism. Eventually, the navigation function of hippocampal formation was successfully simulated to guide eye movement after the weight relations were built between image features (e.g., the nose of a given face) and place cells (termed as feature label cells), the place cells and grid cells, and then grid cells and displacement vector using Hebbian associations. In Banino's study [27], a recurrent network was trained to perform vector-based navigation in a 2.2 m $\times$ 2.2 m square arena. As a result, the network behaves like a mammalian with accurate performance, implicating the great capacity of grid cell population in coding both non-spatial and spatial knowledge.

**5.5 Conclusion: The reference frame system**

By reviewing the latest discoveries of relational memory, we advocate future studies to pay more attention to the reference frame system of the brain. Although some aspects of reference frames, such as the role of populational coding, are already revealed, there is still more work needed to fully understand the nature of reference frames, for example, the diversity, the plasticity, or the brain area-dependent functions of reference frames. Moreover, to our best knowledge, the study of reference frames would also be valuable to promote the transformation of brain-inspired discoveries into AI research in at least following two directions. One is the development of graph theory, where reference frames may be used to solve the subgraph isomorphism problem while the winner-take-all mechanism may help to interpret the routing problem. More important, populational coding would contribute to the biology-based understanding of the concept of nodes and edges. The other is cognitive reasoning and judgment, through which our brain not only retrieves stored knowledge (i.e., discriminative models) but also generates new knowledge (i.e., generative models) beyond experiences. For example, we could use the ability of "imagination" to simulate a future plan repeatedly until the optimized solution is found. To sum up, future studies on reference frames would be helpful for flexible knowledge arrangement, which is the key to reveal the mechanism of relational memory and developing the universal knowledge coding framework for AI.

# Chapter 6 Neural Plasticity: Lessons from Perceptual Learning

*Luyao Chen, Xizi Gong, Fang Fang\**

## 6.1 Introduction: Perceptual learning and plasticity

The brain is a vastly ordered dynamic neural network system involved in almost all vital life activities. Experience can induce changes in the functional properties of neurons and neural circuits in this neural network system, which is referred to as brain plasticity and helps individuals adapt to the environment. Perceptual learning is a typical phenomenon in which the perceptual system adapts to the external environment and refers to the sustained and solid changes in the perception of physical stimuli through repeated practice or experience [1,2,3]. It manifests as a gradual unconscious improvement in the ability to recognize perceptual features and objects maintained steadily over months and years [4]. This increase in perceptual ability is accompanied by neural changes at multiple structural and functional levels of the brain, thus providing an excellent paradigm for studying brain plasticity.

## 6.2 Specificity and transfer of perceptual learning

### 6.2.1 Phenomenology

The traditional view is that the perceptual system is highly plastic during the early stages of individual development. However, this does not mean that brain function is solidified in adult individuals. Perceptual learning, which refers to the long-term performance improvement resulting from practice, has been widely used as a paradigm to study experience-dependent brain plasticity in adults [1,2,3]. Perceptual learning is a typical phenomenon in which the perceptual system adapts to the external environment. It refers to the perception of certain stimuli that produces more sustained and solid changes through practice or experience. Unlike learning in the general sense, perceptual learning does not acquire explicit knowledge but is associated with implicit memory and is expressed as an increase in the ability to discriminate or recognize perceptual features and objects. Perceptual learning can occur in multiple modalities, such as vision, hearing, smell, touch, and taste. Here in this chapter, our review focused on visual perceptual learning.

Specificity and transfer of perceptual learning are defined as whether performance on a second task shows improvement. In early studies, the specificity of visual perceptual learning was demonstrated by the coupling of the acquired behavioral improvement to the physical properties of the trained stimulus and task. It implies that perceptual learning may occur in the low-level cortical areas. The sensory information used in learning originates from the early retinotopic visual areas with small receptive fields [5]. The "early" theory attributes perceptual learning effects to neural tuning changes among neurons in the primary visual cortex [6,7]. Dosher and Lu assort specificity into five kinds according to the involved sites or cortical levels: retinal location, the eye of training, stimulus feature or object, nature of the judgment, and testing context [8]. For each category, partial specificity and partial transfer are common patterns that have accumulated evidence in different observation tasks. Taking the iconic retinal location specificity as an example, Schoups et al. precisely assessed the retinal location specificity in an orientation discrimination task [9]. Participants practiced first at the fovea and then in a series of locations in the periphery around 5° annuli. Results indicate that the learning improvement is specific to locations in the peripheral. Since pre-trained performance in different peripheral locations is better than that at the fovea, it can be concluded that transfer occurs from the fovea to the peripheral locations.

On the other hand, a series of studies indicate that perceptual learning specificity could be reduced or even completely abolished [10,11,12,13,14]. For example, with a double-training paradigm that employed contrast training at one location and orientation training at a second location, Xiao et al. showed a complete transfer of contrast to the second location [11]. Such transfer of perceptual learning indicates that the learning process may also involve the high-level cortex so that more complex processing participants for the different stimuli. In the following part, we will review the single-unit and fMRI studies towards the mechanism of perceptual learning, revealing that learning-related changes occur at multiple stages in the brain, which in turn reconciles the conflict between specificity and transfer of perceptual learning.

**6.2.2 Neural mechanism of perceptual learning**

Corresponding to the specificity of perceptual learning, numerous electrophysiological and fMRI studies have revealed that perceptual learning correlates with enhanced activity in early visual cortical areas. In 2001, Schoups' study on rhesus monkeys linked improved neuronal performance of trained neurons with behavioral improvement in orientation identification [7]. A specific and efficient increase in the slope of neuronal tuning curves was observed for the trained neurons most likely to code the identified orientation. At the same time, no change in the tuning curve was observed for the untrained orientations. In 2010, Hua et al. combined behavioral assessment and extracellular single-unit recording on cats to examine the effects of training in an orientation identification task [15]. Results show that cats' perceptual contrast sensitivity to gratings was significantly improved by training, and the effect was more substantial in the trained eye. For the single-unit recording results in the V1, the neurons' mean contrast sensitivity of the trained cats was significantly higher than that of the untrained cats. Their results suggest that the neuronal contrast gain in V1 induced by training reveals behavioral improvements in perceptual contrast sensitivity. In order to examine the activation changes in the human visual cortex across perceptual learning, Yotsumoto et al. used a texture discrimination task (TDT) in an fMRI study [16]. Results demonstrated that the corresponding BOLD activation in V1 increased with participants' behavioral performance, while it decreased when the performance saturated. Furthermore, evidence from EEG experiments on humans showed a quick rise of the C1 component in the early visual cortex when training improved behavioral performance [17]. It suggests that the increase of early visual area response by perceptual learning results from local receptive field changes, not feedback from later visual areas. In addition, an MRI study applying neurofeedback techniques improved behavioral performance by removing external visual input and relying on training neural signals from the visual cortex alone, supporting the idea that the primary visual cortex plays a decisive role in perceptual learning [18]. In 2016, Yu et al. used a contrast detection task in a 30- days training on a specific eye and visual hemifield to investigate perceptual learning in subcortical nuclei [19]. Results showed that the fMRI signal of the subcortical structures, the lateral geniculate nucleus (LGN), increased to low-contrast patterns. The increase was specific for the trained eye and visual hemifield and only occurred in the magnocellular layers, while it was absent in the parvocellular layers of LGN even when subjects did not pay attention to the patterns. These findings suggest that neural plasticity induced by perceptual learning might occur as early as at the thalamic level.

However, it has also been identified that brain areas related to attention and decision-making are associated with perceptual learning [2,20,21,22]. Specifically, perceptual learning was found to be related to enhanced selectivity or attenuated neural responses in the frontoparietal areas involved in the intraparietal sulcus (IPS), the external parietal lobe (LIP), and the anterior cingulate gyrus (ACC) [23,24,25]. Moreover, the reverse hierarchy theory claims that it is the top-down rather than the bottom-up information flow that governs the learning process [26]. Learning first occurs at the task-specific high-level layer, then is achieved at lower-level layers if necessary. This theory, to a certain extent, explained the occurrence of transfer in perceptual learning. Meanwhile, some studies also indicated that perceptual learning is associated with increased functional connectivity between visual and decision-making areas [27,28]. The reweighting theory suggests that perceptual learning does not alter the functional properties of the early visual cortex; instead, it changes the strength of the connections (weights) between neurons that represent visual information and decision units [29]. Based on the reweighting theory, Law and Gold modeled learning as a process that a high-level decision neuron refined its connection weight with the sensory neurons specifically for a trained motion direction [24,25,26,27,28]. A perceptual learning study on motion discrimination demonstrated that behavioral improvement could be explained by the changes in the sensory representation of stimuli in V3A as well as in the connectivity from V3A to IPS [27]. Similarly, to investigate the way perceptual learning modulates the activity of decision-related areas in the human brain, a model-based approach was adopted with a motion direction discrimination task [30]. Besides the enhanced neural responses in the frontoparietal network and the decision network, results showed strengthened connectivity from V3A to the ventral premotor cortex (PMv) and from IPS to frontal eye field (FEF) was paralleled with training. In all, the mechanism of perceptual learning is not a simple process but a complex interaction among multiple brain areas, such that both specificity and transfer could be observed under different conditions.

### 6.2.3 The role of GABA in perceptual learning

GABA is a molecule involved in the brain's inhibitory modulation of neurons. It has been demonstrated in previous animal studies that GABAergic inhibition plays a significant role in learning and synaptic plasticity [31,32]. Moreover, human MRS studies have revealed that the GABA concentrations in the visual cortex are associated with homeostatic plasticity [33], while those in the motor cortex are associated with individual abilities [34,35] and improved performance in motor learning [36,37,38]. Specifically, in visual perceptual learning, the effect of GABA level on the performance improvement of two visual tasks showed opposite directions [39]. In the target-detection task, subjects did better if their GABA decreased after training, while in the feature-discrimination task, subjects did better if their GABA increased after training. Evidence from pharmacological manipulations also highlights the significant role of GABA in learning. By chronically using the selective serotonin reuptake inhibitor (SSRI) fluoxetine, adult rats' visual cortex plasticity could be strengthened, while the strengthening effect would be blocked if diazepam was used to increase GABA [40].

### 6.3 Hebbian rule and the computational models for perceptual learning

In 1949, Hebb stated that "when an axon of cell A is near enough to excite a cell B and repeatedly or persistently takes part in firing it, some growth process or metabolic change takes place in one or both cells such that A's efficiency, as one of the cells firing B, is increased" [41]. Long-term potentiation (LTP) and long-term depression (LTD), which were induced by high- and low-frequency intermittent stimulation, respectively, jointly allow bidirectional synaptic modification and are believed to be the synaptic basis of learning and memory [42,43,44]. Further, spike timing–dependent plasticity (STDP) indicates that the direction of synaptic modification depends on the temporal order of pre and postsynaptic spiking [45,46,47,48,49]. Specifically, repeated exposure to stimuli affects the temporal relationship between pre- and postsynaptic activation, then the synapse strength changes [50]. The changes may reduce the synaptic latency [51] and sharpen the timing of neural activations.

It is suggested that LTP may underlie perceptual learning. A. Sales et al. provided direct evidence from electrophysiological recordings in slices that perceptual learning influenced the LTP of synaptic responses in rat's V1 [52]. As is known, it is challenging to study synaptic modifications in vivo. It deserves to be mentioned that in a human behavioral study, LTP-like passive high-frequency stimulation improved subjects' performance in the luminance change detection task, whereas LTD-like passive low-frequency stimulation impaired the performance [53]. Since it is experimentally challenging to investigate whether the duction of LTP could induce perceptual learning, many studies show similarities between visual perceptual learning and LTP and the role of visually evoked potentials (VEPs) as intermediates [54]. For example, as mentioned above, the specificity of perceptual learning coupled the behavioral improvement with the trained features. Similarly, LTP is specific to the activated synapses though the cell globally produced the late-LTP required proteins [54].

Based on the phenomenon and mechanism of perceptual learning, the reweighting theory was proposed to explain and predict experimental results. This theory claims that the primary loci of perceptual learning are at a more integrated level that correlates with the task-relevant decision [55]. Perceptual learning changes the weighting of representations in the visual system at different levels. The principles are to strengthen the connections from the tuned visual channels to the learned categorization structure and to reduce inputs from task-irrelevant channels. It has been demonstrated that the corresponding augmented Hebbian reweighting model (AHRM) based on this theory can generate an extensive range of performance patterns that are comparable to empirical data [56].

Modeling studies also indicate the importance of GABA in perceptual learning. In 2011, with a cortical neural network model, Osamu examined the role of ambient GABA in leaving memory traces for perceptual learning [57]. Further, with a novel neuron-astrocyte network model, Osamu studied the effect of astrocytic gap-junctional communication on perceptual learning [58]. Results showed that the synchronization of local ambient GABA enhanced the STDP level, which facilitated perceptual learning by influencing the process of leaving memory trace during learning.

**6.4 Conclusion: New tech and new insight**

Learning in the human brain is a complex and flexible process involving both the Hebbian rule-based synaptic strength changes and high-level scheduling of interactions between multiple brain regions. Ideally, we would like to comprehensively observe the neural activity behind a specific behavior (for example, learning) with high temporal-spatial resolution. Though current technologies are challenging to satisfy the demand for concurrent high temporal and spatial resolution, we can still take a glimpse at the intrinsic key between learning and neuroplasticity to form a comprehensive view with the help of single-neuron recording, EEG/EMG recording, fMRI acquisition, and the computational models. The Hebbian rule has been demonstrated in various neural circuits from insects to humans and has laid the foundation for constructing neural network models with learning functions. It is known that the connection weights between nodes in current artificial neural networks are usually fixed after training, which makes it different from human learning in terms of transfer and functional compensation. In human perceptual learning, transfer and specificity are balanced by an unclear mechanism. The balance between transfer and specificity makes the brain an efficient and energy-saving intelligence center. Future research may collect direct evidence from the human brain to reveal the synaptic-level mechanism of learning. By benefiting from human studies, artificial neural networks may evolve more robust learning strategies.

**Conclusion: Inter-disciplinary research to unlock the nature of intelligence**

In this survey, we have summarized six first principles which we believe are fundamental to the intelligence of the brain, and they are likely inspirational to future development of AI. These six principles are:

- In Chapter 1, we introduced **attractor neural networks**, which are canonical network models for neural information processing. These attractor networks have appealing computational properties that endow the brain with the capacities of representing information robustly, searching information efficiently, and integrating information optimally. They may serve as the building block to develop brain-inspired general intelligence.

- In Chapter 2, we introduced the concept of **criticality**, referring to a special but general state of a dynamical system. The experimental data have suggested that the brain works in a critical state, which enables the brain to process information rapidly and efficiently and to hold many other computational advantages. Recent research has begun to explore the potential gains of being at the critical state in reservoir and deep neural networks.

- In Chapter 3, we introduced **random networks**. Countering the common sense that functionality arises from structured networks, the brain networks exhibit large randomness in neuronal connectivity. Research has shown that these random networks can actually have many computational advantages, such as being universal function approximators, physically easy to implement, saving training costs, expanding to high-dimensional representation space, and implementing locality-sensitive hashing algorithms effectively. The idea of random networks is also attracting the attention of AI researchers.

- In Chapter 4, we introduced **spare coding**, a strategy referring to the observation that the brain utilizes a small number of neurons or spare neuronal responses to represent information. Sparse coding has been widely observed in the experimental data and is believed to have many computational advantages, such as representing information efficiently, simplifying the decoding of information, and effectively achieving compressed sensing.

- In Chapter 5, we introduced the concept of **relational memory**. Recent experimental studies have indicated that the brain utilizes the reference frame to store the relations between memories, including both spatial and non-spatial memories. This occurs in the MTL, and grid cells are the neural substrate to implement the reference frame. Therefore, understanding how relational memory is realized in the brain will help us eventually build conceptual knowledge in AI.

- In Chapter 6, we introduced **neural plasticity**, the neural substrate for the brain to acquire new experiences and form new knowledge. Perceptual learning is one approach to achieve neural plasticity. Depending on the learning task or protocol, the learning performance can be either location- or feature-specific, or transferable among locations or features, and the learning mechanism involved can be either Hebbian or high-level connection re-weighting. Therefore, understanding how the brain is modified through learning may shed light on the development of self-evolution of ANN's structure.

In summary, we hope this survey would be valuable to AI researchers, and the fundamental principles we have introduced can be inspirational to develop more advanced AI architecture, algorithms, and systems so that AI can achieve human-like intelligence.

**Acknowledgement**

This work is an outgrowth of a series of talks presented and discussed at the internal ABC seminar (AI of Brain and Cognitive Sciences) at BAAI, with the purpose of promoting the interdisciplinary research among neuroscience, cognitive science, computational science and artificial intelligence, and exploring the new methods of brain-inspired next generation artificial intelligence. We would like to especially thank Ms. Yaqiong Yan for organizing this work. The work is supported by Beijing Academy of Artificial Intelligence.

**Chapter 2**

[43] Lukoševičius M and Jaeger H. Reservoir computing approaches to recurrent neural network training. *Computer Science Review*, 2009; 3: 127-149 (2009).

[44] Lukoševičius M, Jaeger H and Schrauwen B. Reservoir computing trends. *KI-Künstliche Intelligenz*, 2012; 26: 365-371.

[45] Zeng G, Huang X, Jiang T *et al*. Short-term synaptic plasticity expands the operational range of long-term synaptic changes in neural networks. *Neural Networks* 2019; 118: 140-147.

[46] Choi J and Kim P. Reservoir computing based on quenched chaos. *Chaos, Solitons & Fractals* 2020; 140: 110131.

[47] Poole B, Lahiri S, Raghu M *et al*. Exponential expressivity in deep neural networks through transient chaos. *Advances in neural information processing systems*, 29 (Barcelona 2016).

[48] Schoenholz SS, Gilmer J, Ganguli S *et al*. Deep information propagation. *International Conference on Learning Representations* (San Juan 2016).

[49] Saxe AM, McClelland JL and Ganguli S. Exact solutions to the nonlinear dynamics of learning in deep linear neural networks. *International Conference on Learning Representations* (Scottsdale 2013).

[50] Pennington J, Schoenholz S and Ganguli S. Resurrecting the sigmoid in deep learning through dynamical isometry: theory and practice. *Advances in neural information processing systems*, 30 (Long Beach 2017).

[51] Pennington J, Schoenholz S and Ganguli S. The emergence of spectral universality in deep networks. *In International Conference on Artificial Intelligence and Statistics* (pp. 1924-1932). PMLR (Lanzarote 2018).

[52] Oprisa D and Toth P. Criticality & Deep Learning I: Generally Weighted Nets[J]. arXiv preprint arXiv:1702.08039, 2017.

[53] Farrell M, Recanatesi S, Moore T *et al*. Gradient-based learning drives robust representations in recurrent neural networks by balancing compression and expansion[J]. *Nature Machine Intelligence* 2022; 4: 564-573.

**Chapter 3**

[1] Hubel DH and Wiesel TN. Receptive fields, binocular interaction and functional architecture in the cat's visual cortex. *The Journal of physiology* 1962; 160(1): 106.

[2] Rigotti M, Barak O, Warden MR *et al*. The importance of mixed selectivity in complex cognitive tasks. *Nature* 2013; 497(7451): 585-590.

[3] Fusi S, Miller EK and Rigotti M. Why neurons mix: high dimensionality for higher cognition. *Current opinion in neurobiology* 2016; 37: 66-74.

[4] Johnston WJ, Palmer SE and Freedman DJ. Nonlinear mixed selectivity supports reliable neural computation. *PLoS computational biology* 2020; 16(2): e1007544.

[5] Barak O, Rigotti M and Fusi S. The sparseness of mixed selectivity neurons controls the generalization–discrimination trade-off. *Journal of Neuroscience* 2013; 33(9): 3844-3856.

[6] Parthasarathy A, Herikstad R, Bong JH *et al*. Mixed selectivity morphs population codes in prefrontal cortex. *Nature neuroscience* 2017; 20(12): 1770-1779.
45

**Chapter 4**

## Chapter 5